\documentstyle[12pt,epsf]{article}
\def\lsim{\mathrel{\rlap {\raise.5ex\hbox{$ < $}}
{\lower.5ex\hbox{$\sim$}}}}

\newcommand{\pr}{\paragraph{}}
\newcommand{\be}{\begin{equation}}
\newcommand{\ee}{\end{equation}}
\newcommand{\bea}{\begin{eqnarray}}
\newcommand{\nn}{\nonumber}
\newcommand{\eea}{\end{eqnarray}}

\def\gappeq{\mathrel{\rlap {\raise.5ex\hbox{$>$}}
{\lower.5ex\hbox{$\sim$}}}}

\def\lappeq{\mathrel{\rlap{\raise.5ex\hbox{$<$}}
{\lower.5ex\hbox{$\sim$}}}}

\begin{document}

\begin{titlepage}
\begin{flushright}
ACT-19/97 \\
CTP-TAMU-50/97 \\
NTUA-68/97\\
OUTP-97-69P \\
UA-NPPS-9-97\\
December  1997\\
gr-qc/9712051 \\
\end{flushright}

\begin{centering} 
\vspace{.05in}
{\large {\bf Irreversible Time Flow 
in a Two-Dimensional Dilaton Black Hole with Matter}} \\
\vspace{.05in}
G.A. Diamandis$^{a,*}$,  John Ellis$^{b} $,
B.C. Georgalas$^{a,*}$
N.E. Mavromatos$^{a}$, \\
D.V. Nanopoulos$^{c}$, and E. Papantonopoulos$^{d}$ \\ 
\vspace{.02in}
\vspace{.1in}
{\bf Abstract} \\
\vspace{.02in}
\end{centering}
{\small  We show that an exact solution 
of two-dimensional dilaton gravity with matter discovered previously
exhibits an irreversible temporal flow
towards flat space with a vanishing cosmological constant.
This time flow is induced by the
back reaction of matter on the
space-time geometry.
We demonstrate that the system is not in equilibrium if
the cosmological constant is non-zero, whereas the
solution with zero cosmological constant is stable. The
flow of the system towards this stable end-point is
derived from the renormalization-group flow of the
Zamolodchikov function. This behaviour is
interpreted in terms of non-critical Liouville string, with the 
Liouville field identified as the target time.}

\vspace{0.02in}

\begin{flushleft}
$^{a}$ University of Oxford, Dept. of Physics
(Theoretical Physics),
1 Keble Road, Oxford OX1 3NP, United Kingdom   \\
$^b$ Theory Division, CERN, CH-1211 Geneva, Switzerland  \\
$^{c}$ Center for
Theoretical Physics, Dept. of Physics,
Texas A \& M University, College Station, TX 77843-4242, USA,
Astroparticle Physics Group, Houston
Advanced Research Center (HARC), The Mitchell Campus,
Woodlands, TX 77381, USA, and
Academy of Athens, Chair of Theoretical Physics, 28 Panepistimiou Ave.,
Athens GR-10679, Greece.\\
$^{d}$ Physics Department, National Technical University, 
Zografou GR-157 80, Athens, Greece, \\
$^{*}$ On leave from 
Nuclear and Particle Physics Section, Physics Department, 
Athens University, 
Panepistimiopolis,
Athens GR-157 71, Greece.\\
\end{flushleft}

\end{titlepage}
\newpage

\section{Introduction}

Two-dimensional black holes~\cite{Witten,Wadia,Verlinde} 
are a useful laboratory for studying fundamental issues 
of quantum gravity, such as the structure 
of black-hole space times and the information paradox.
One of their most useful features is their exact solubility.
Moreover, despite their two-dimensional nature, these black holes exhibit
many of the interesting and non-trivial features of their 
higher-dimensional counterparts, namely 
horizons related to the exponential of the dilaton field, 
Hawking radiation~\cite{Callan,Russo}, 
non-thermal back reaction effects~\cite{radiation}, etc..
Recently, many of the features found previously in
two-dimensional stringy black-hole models have been recovered 
in the $D$-brane description of black holes in higher-dimensional
space times.
\pr
The two-dimensional stringy black hole has also served as a
prototype system for developing the interpretation of 
the world-sheet Liouville field as a target-space temporal evolution 
parameter~\cite{emn,kogan,emncosmol}. 
The propagating light matter modes~\footnote{Confusingly called `tachyons',
though they are in fact massless.} 
of the two-dimensional
stringy black hole exhibit non-trivial 
conformal dynamics due to their coupling
with discrete quasi-topological higher-level modes, which
constitute an `environment' that is not observable in conventional
low-energy scattering experiments~\cite{emn}. These modes are
therefore integrated out in the effective low-energy theory
of the light modes,
which exhibits coherence loss due to this entanglement with
the higher-level modes. It has been argued~\cite{EMND} on the basis of a
general
renormalization-group analysis and explicit $D$-brane
examples that these features are shared by more realistic
string models in higher dimensions.
\pr
In this paper we investigate these arguments using
an exact solution of two-dimensional dilaton gravity with matter
which some of us have discovered previously. We argue here that it
presents features suggested in the
approach of~\cite{emn,emncosmol}. In particular,
the Liouville field may be identified as the target time-evolution
parameter in the two-dimensional
dilaton-gravity-matter system. 
This two-dimensional black-hole space time features infalling
matter and a
non-zero cosmological constant. It is unstable, and is interpreted as an
time-dependent non-equilibrium
black hole, which exhibits a potential flow
towards a flat space time with a zero comsological constant,
which is stable.
\pr
Our first demonstration of the instability of the 
solution with non-zero cosmological solution is made by
evaluating the coefficients of the Bogolubov transformation
between the asymptotic `in' and `out' states.
These yield
particle creation with a number density that deviates from a
thermal equilibrium distribution, if the cosmological constant is
non-zero. On the other hand, there is no particle production in
the limit of zero cosmological constant. We then show how these
results can be interpreted in the context of Zamolodchikov's
$c$ theorem as flow towards a fixed point of the renormalization
group, exhibiting explicitly the rate of this flow in the
weak-field approximation. This analysis is based on the
identification~\cite{emn,emncosmol,kogan} of the (world-sheet) 
renormalization
scale, i.e., the renormalization-group evolution parameter, with
the world-sheet Liouville field and the target time variable.

\section{Review of the Two-Dimensional Dilaton Black Hole with Matter}

For completeness, we first review relevant aspects
of the dilaton-matter black-hole solution found in~\cite{tachyon}. 
In two dimensions, the action of the dilaton-tachyon system coupled to
gravity is
 \begin{equation}
 S=\frac{1}{2\pi}
 \int d^2x\sqrt{-g}
 \{e^{-2\phi}
 [R+4(\nabla\phi)^2-(\nabla T)^2
 -V(T)+4\lambda^2 ] \}
 \label{action}
 \end{equation}
where $4\lambda ^2$ is the two-dimensional cosmological constant,
which is related to the central charge $c$ of the corresponding
world-sheet $\sigma$ model via~\cite{Witten}: 
\be
    \lambda^2=\frac{k}{3}(c-26) 
\label{lamb}
\ee
where $k >2$ is the level parameter 
of the Wess-Zumino conformal field theory that has as a 
conformal solution 
the target-space two-dimensional theory (\ref{action}), 
which is in turn related to the central charge by~\cite{Witten} 
$c= (3k / k-2)-1$. The quantity 
$V(T)$ is the tachyon potential, which is ambiguous
in string theory~\cite{Banks}. 
The only unambiguous term is 
the quadratic term for the tachyon field, which 
is also all that we need for our analysis.  The equations of
motion resulting from the action (\ref{action}),
which are equivalent at ${\cal O}(\alpha ')$
to the conformal-invariance conditions of the $\sigma$ model,
are
 \bea
&~& \Box\phi -(\nabla\phi)^2+\frac{1}{4}R
 -\frac{1}{4}(\nabla T)^2
 +\frac{1}{2} T^2+\lambda^2 =0 \nn \\
&~& \Box T-2(\nabla\phi)(\nabla T)+2T=0 \nn \\
&~& \Box\phi-2(\nabla\phi)^2
 +T^2+2\lambda^2=0 
\label{equation}
\eea
These equations can be analyzed most easily in the
conformal gauge
      \[g_{\mu\nu}=e^{2\rho}n_{\mu\nu}
      \quad ,
      \quad
      n_{\mu\nu}=diag(-1,1) \]
and if we use light-cone coordinates 
$x^\pm = x^0\pm x^1$. From now on, we shall refer to $\rho$ 
as the conformal factor, and we shall follow~\cite{emn} by
identifying it with the Liouville field.
\pr
An exact solution of this model with a
non-trivial time-dependent tachyon configuration 
was found in~\cite{tachyon},
by assuming that the tachyon $T$ is a function of $x^+$.~\footnote{We
note that this is a consistent possibility because 
the tachyon field in a two-dimensional target-space time may be written
with canonical kinetic terms if one rescales
with the dilaton field:
${\tilde T} = e^{-\Phi} T$. This means that $T$ can still be a function of
$x^{+}$ even if
the dilaton is in general a function of $x^{\pm}$. Moreover, this is 
consistent with a mass term appearing in the equation of 
motion for ${\tilde T}$.}
The following solution to the system of equations (\ref{equation})
was obtained in~\cite{tachyon}:
     \begin{equation}
     e^{-2\phi}=e^{-2\rho}=
     C_1 F(1, \frac{\lambda^2}{2}+1 ; -\frac{T^2}{4}) +
     C_2 (\frac{T^2}{4})^{-\frac{\lambda^2}{2}} e^{- \frac{T^2}{4}}
     -\frac{2 e^{-\frac{T^2}{4}}}
      {a (T^2)^{\frac{\lambda^2}{2}}} x^-
      \label{fullsol}
      \end{equation}
where $F(\alpha ,\beta ; c)$ is the confluent hypergeometric 
function, and $a > 0$, 
$C_1 > 0$ and $C_2$ are integration constants. 
The solution has been expressed in terms of the 
tachyon field, which satisfies:
       \begin{equation}
       \frac{d T^2}{d x^+}=
       -a (T^2)^{\frac{\lambda^2}{2} +1} e^{\frac{T^2}{4}}
       \label{eq:two1}
       \end{equation}
The parametrization 
in terms of the tachyon field was made
possible by the monotonic behaviour of the field $T$, which
admits the physical interpretation of the tachyon field $T$ as an 
infalling matter field. The relation (\ref{eq:two1}) 
can be integrated in a closed form to give:
\be
x^+=\frac{1}{4^{\frac{\lambda ^2}{2}}a}
\Gamma (-\frac{\lambda ^2}{2},\frac{T^2}{4})
\label{completesolution}
\ee
where $\Gamma (\alpha, \beta )$ denotes the incomplete Gamma 
function~\cite{erdelyi}. The asymptotic behaviour of the matter 
field is such that at $x^{+} \rightarrow \infty (0)$ the matter
background $T^2 \rightarrow 0 (\infty)$.  
For later use, we also present here an expression 
for the curvature scalar,
which in two dimensions determines completely the Riemann tensor:
\be
 R=\frac{4\lambda ^2C_1}{C_1 F(1, \frac{\lambda^2}{2}+1 ; -\frac{T^2}{4})
+ C_2 (\frac{T^2}{4})^{-\frac{\lambda^2}{2}} e^{- \frac{T^2}{4}}
 -\frac{2 e^{-\frac{T^2}{4}}}{a (T^2)^{\frac{\lambda^2}{2}}} x^-}
\label{curvature}
\ee
We see immediately that in this model one obtains a flat space time
if the cosmological constant $\lambda^2$ vanishes, provided 
one stays away from the initial singularity at $x^+=0$, where the 
tachyon field diverges.~\footnote{Note 
that the curvature scalar
becomes $R \simeq 2 T^2 $ at the singularity, 
and thus remains singular, whatever the value of
$\lambda$. In the specific case $\lambda =0$, the curvature scalar 
has the form $ \delta (x^+)$. When one
avoids the initial singularity using a cut-off, 
the space is completely flat in the case $\lambda =0$.}
\pr
For general $\lambda ^2 \ne 0 $, i.e., central charges 
$c \ne 26$, as seen in (\ref{lamb}), 
the above solution has the structure of 
a non-static time-dependent black hole,
with the important feature~\cite{tachyon} 
that there is a tachyon singularity at $x^{+}=0$,
which is displayed in Fig.~1. Thus
there is no white hole in this model, 
which we find suggestive that the model may exhibit an arrow of time.

\begin{centering}
\begin{figure}[t]
\epsfxsize=4in
\centerline{\epsffile{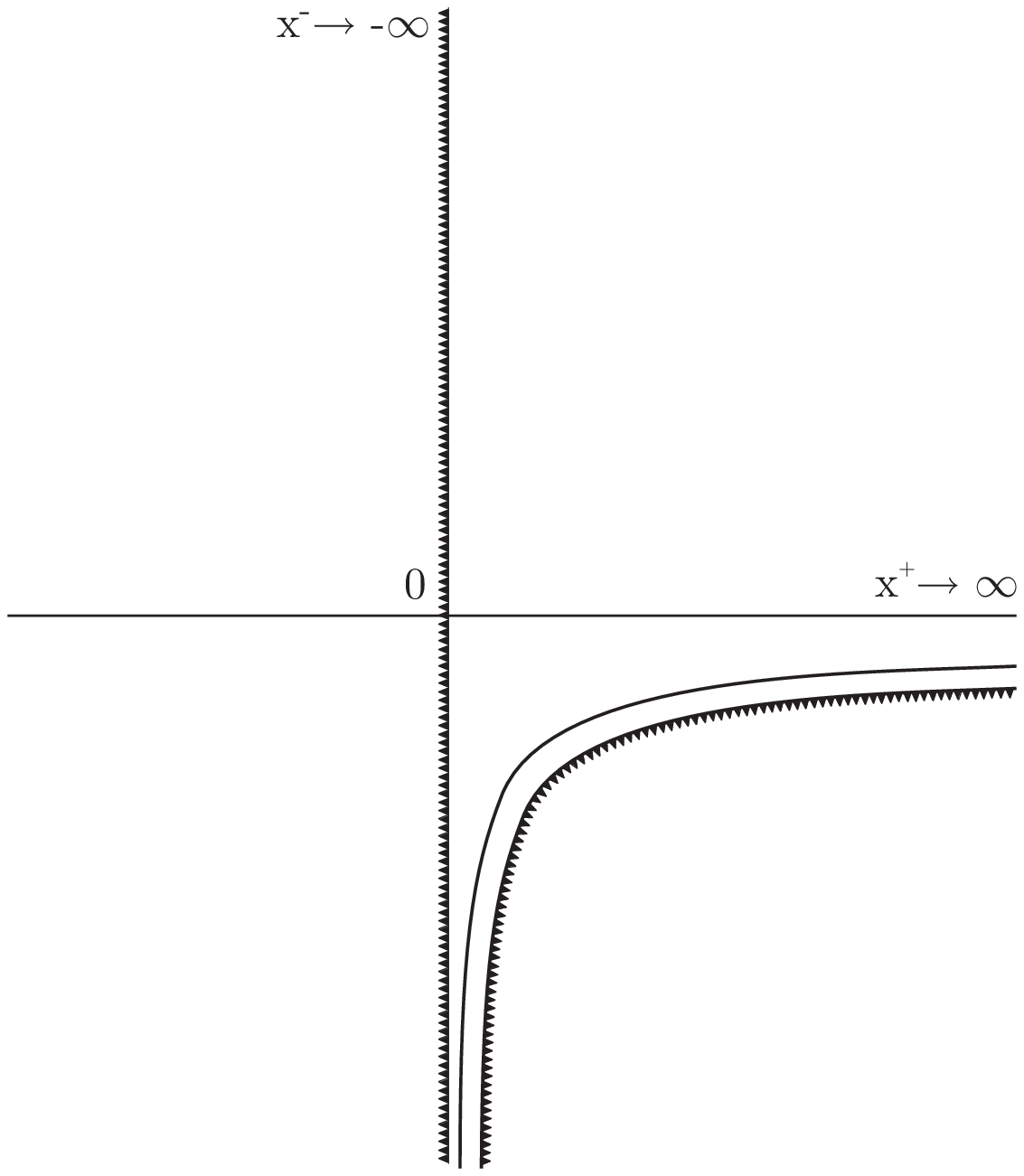}}
\vspace{1cm}
\caption{{\it The non-static 
solution of the equations of motion 
corresponding to an infalling matter 
$T=T(x^+)$. The singularities are indicated 
by jagged lines, and the apparent horizon is
represented by the thin continuous line. It is clear that there is no 
white hole.}}
\label{fig1}
\end{figure}
\end{centering}

\section{Analysis of Stability and Particle Production}

Motivated by this indication of a possible
arrow of time in this model, we
now look for a connection with the irreversible 
Liouville time flow along the lines suggested in~\cite{emn,kogan}. 
The first step in this search is to find a coordinate
transformation which makes
the Liouville mode, i.e., the conformal factor $\rho$ 
in the metric (\ref{fullsol}), linear 
in the transformed time 
variable. As we discuss below, 
this is possible only when 
the cosmological constant $\lambda^2=0$, 
corresponding (\ref{lamb}) to a
critical value for the central charge, and representing
a stable, equilibrium configuration. 
For the other values of $\lambda ^2 > 0$, the conformal factor also
has a spatial dependence, as we also discuss below. 

\subsection{The $\lambda = 0$ Case: Stability}

We first examine the structure of the solution at the point 
$\lambda ^2 =0$, corresponding to $c=26$, where the space time is flat
(\ref{curvature}). In this case, the metric simplifies to:
\be
   e^{-2\phi}= e^{-2\rho} =[ (C_1 + C_2) - \frac{2}{a}x^{-}]e^{-\frac{T^2}{4}}
\label{metrzero}
\ee
Setting $C_1=-C_2$ by making a shift in $x^{-}$,
the metric corresponding to (\ref{metrzero}) becomes:
\be
ds^2 = \frac{a}{2} \frac{e^{\frac{T^2}{4}}}{x^{-}} dx^{+}dx^{-} 
\label{metricelementzero}
\ee
where the tachyon field $T$ obeys the equation
\be
\frac{dT^2}{dx^{+}}=-a T^2 e^{T^2/4}
\label{tachyonequ}
\ee
We can use (\ref{tachyonequ}) to recast (\ref{metricelementzero})
in the form:
\be
 ds^2 = -\frac{1}{2}\frac{dT^2 dx^{-}}{T^2x^{-}}
\label{metrtach}
\ee
which is the basis for further discussion of the structure of this
solution.
\pr
In order to establish a connection between the dilaton and the
target time variable, we first consider
the coordinate transformation $x^{\pm} \rightarrow (r',t')$
with
\be 
   r'={\rm ln}(-y^+y^-) , \qquad t'=-{\rm ln}(-\frac{y^-}{y^+})      
\label{transform}
\ee
where $y^{\pm}$ are the corresponding 
Kruskal-Szekeres coordinates: 
\be
 T^2 = \frac{1}{y^+}, \qquad , \qquad x^-=y^-
\label{xpmy}
\ee
In this coordinate system, 
the matter background (\ref{xpmy}) is 
singular at 
$y^+=0$, whilst the  
metric element becomes that of Minkowski space time. 
The background dilaton has the form
\be
  \phi = \frac{1}{2}{\rm ln}(a/2) + \frac{t'-r'}{4} 
+ \frac{1}{8} e^{-(r'+t')/2}
\label{dilatonback2}
\ee
which has the asymptotic behaviour: $\phi \rightarrow -r'/4$, as 
$r' \rightarrow \infty$ for $t' $ finite, 
and $\phi \rightarrow t'/4$ as $t' \rightarrow
\infty$ for $r'$ finite. Thus
the dilaton may have a linear asymptotic dependence on
time~\cite{aben,emn},
in the presence of non-trivial matter. 
\pr
The above representation is not convenient for deriving
tractable closed-form expressions, for which purpose 
it is better to consider an
alternative coordinate transformation $x^{\pm}
\rightarrow (r,t)$ where
\be 
   r={\rm ln}(-{\tilde x}^+ {\tilde x}^-) , \qquad 
t=-{\rm ln}(-\frac{{\tilde x}^-}{{\tilde x}^+})      
\label{transform2}
\ee
so that ${\tilde x}^{\pm}$ are a new pair of
Kruskal-Szekeres coordinates. We define ${\tilde x}^- \equiv x^-$
and
\be
({\tilde x}^+)^2 \equiv - {\rm ln} ({T^2 \over \beta})
\label{xpmy2}
\ee
The validity of this coordinate transformation follows from the
monotonic behaviour of the tachyon solution. We shall use
(\ref{xpmy2}) as a coordinate patch when $T^2 \le \beta$, i.e.,
when ${\tilde x}^+ \ge 0$. The parameter $\beta$ can be
regarded as `brick-wall' cut-off that shields the matter singularity that
appears when $y^+ = 0$ in the previous coordinate system.
After the transformation (\ref{xpmy2}), the  
metric element becomes:
\be
ds^2=-\left(-\frac{{\tilde x}^+}{{\tilde x }^-}\right)d{\tilde x}^+
d{\tilde x}^- =\frac{1}{4}e^{(r + t)} (-dt^2 + dr^2) 
\label{metrtilda}
\ee
and the conformal factor $\rho$ 
becomes a simple linear function of time:
\be
        \rho = \frac{t}{2} 
\label{rhoconf}
\ee
The dilaton background, however, still depends on both $(r,t)$:
\be
  \phi = \frac{1}{2}{\rm ln}(a/2) + \frac{t-r}{4} 
+ \frac{\beta}{8}{\rm exp}[-e^{r+t}]
\label{dilatonback3}
\ee
with the linear asymptotic behaviours: $\phi \rightarrow 
-r/4$, as $r \rightarrow \infty$ for $t $ finite, and $\phi \rightarrow
t/4$ as $t \rightarrow
\infty$ for $r$ finite. 
\pr
The physical interpretation of the metric 
(\ref{metrtilda}) becomes transparent  
in terms of the following variables:
\be
     \xi \equiv e^{r/2} \qquad ; \qquad \tau = t/2   
\label{xi}
\ee
Looking at the metric element as a function of ($\xi, \tau $):
\be
   ds^2= e^{2\tau}\left( -\xi^2 d\tau ^2 + d\xi ^2 \right) 
\label{rindler} 
\ee
we see that the space is conformally equivalent to a two-dimensional 
Rindler space with constant acceleration~\cite{birrel}.~\footnote{In
our normalization, the acceleration is unity.} We
notice also the appearance of a 
conformal factor $e^{2\tau }=e^{t}$ corresponding to an
expanding universe. This suggests an alternative
representation in terms of the following variables:
\be
   \eta \equiv e^{t/2} \qquad; \qquad r'=r/2 
\label{milnetrans}
\ee
which maps (\ref{rindler}) into a 
two-dimensional Universe that is conformally equivalent to a Milne 
space-time~\cite{birrel},
but with a conformal factor that depends only on space:
\be
   ds^2= e^{2r'}\left( -d \eta^2 + \eta^2 dr ^{'2} \right) 
\label{milne} 
\ee
It is known~\cite{birrel} that both Rindler and
Milne spaces exhibit particle creation, related to the 
cosmological expansion. The peculiarity of our 
metric (\ref{rindler}) or equivalently (\ref{milne}) resides 
in the fact that 
it constitutes a mixing of the two, and hence one has to 
reconsider the analysis of particle production for this case. 
\pr
For this purpose, we now compute the Bogolubov coefficients 
corresponding to the metric (\ref{metrtilda}). This may
conveniently be achieved by the following coordinate
transformation:
\bea 
&~& \xi' \equiv \frac{1}{4}e^{2(r+t)} + \frac{1}{4} (r-t) \nn \\
&~& \eta' \equiv \frac{1}{4}e^{2(r+t)} - \frac{1}{4} (r-t) 
\label{change}
\eea
where $\eta'$ may be regarded as a time variable.
The space parametrized by these $(\xi', \eta')$ variables is considered 
as the `out' space, with the `in' space characterized by the metric
(\ref{metrtilda}). 
The Bogolubov transformations~\cite{birrel} are given in the usual way by: 
\bea
&~&  \alpha = <\Phi ^{out} | \Phi ^{in} > =
\int dr \left( u^{in*} \partial _t u^{out} - \partial _t u^{in*} u^{out}
\right)  \nn \\
&~& |\alpha |^2 - |\beta |^2 = 1
\label{bogol} 
\eea
where 
\be 
   \Phi^{out} = \alpha \Phi ^{in} + \beta \Phi ^{out*} 
\label{defi}
\ee
and the $\beta$ coefficient is associated with particle creation. 
It is straightforward to see that in our case: 
\be 
 \alpha = 1 \qquad ; \qquad \beta =0
\label{10}
\ee
independently of the mode energies and momenta.
Thus, there is {\it no} particle production in this metric.
\pr 
This means that the model with $\lambda = 0$ is a stable
equilibrium state. Perhaps surprisingly,
the effects of the spatial expansion visible in (\ref{rindler})
and the Rindlerlike acceleration~\cite{unruh} cancel
exactly, as far as particle production is concerned.

\subsection{The $\lambda \ne 0$ Case: Instability} 

The same is not the case in
the case $\lambda^2 \ne 0$, as we discuss now.
{}From the point of view of the 
target-space theory (\ref{action}), this case corresponds to a
time-dependent black-hole solution~\cite{tachyon}.
We would therefore expect that it
corresponds to a non-equilibrium situation, since
the non-trivial time dependence means that
a temperature cannot be defined for the system.
To demonstrate this explicitly, we compute the 
Bogolubov coefficients of the vacuum described by the 
$\lambda \ne 0$ model, and then the corresponding particle creation number
seen by an observer who initially observes a flat space time 
at $t \rightarrow -\infty$.
\pr 
The line element in this case is given by (\ref{fullsol}). 
For simplicity, we consider the case with  
weak matter fields $|T| << 1$, such that 
\be
\frac{\lambda ^2}{2}|{\rm ln}(\frac{T^2}{4})| << 1
\label{cond}
\ee
For convenience, and without loss of generality, 
we take $C_1 = -C_2$. The line element 
can then be approximated by: 
\be
ds^2=\frac{\frac{dT^2}{T^2}dx^-}{a4^{\frac{\lambda ^2}{4}}C_1~\frac{\lambda ^2}{2}~{\rm ln}\frac{T^2}{4} - 2x^{-}}
\label{dslightcone}
\ee
in light-cone coordinates, where we have used (\ref{eq:two1}). 
To facilitate our analysis, we assume both an infrared cut-off 
$\Lambda$ and an ultraviolet cut-off $\epsilon$
in time. The matter field $T$, 
defined in (\ref{eq:two1}), acquires a minimum value 
$T^2(\Lambda) = T^2_{min}$, 
such that (\ref{cond}) is satisfied for $T=T_{min}$,
and a maximum value $T^2_{max}-T^2_{min} \equiv 4 {\cal B} > 0$, 
where ${\cal B}$ a constant related to the cut-offs.
\pr
Next we define a convenient
coordinate basis ${\tilde \sigma}^+$, ${\tilde \sigma}^{-}$ by:
\be
e^{{\tilde \sigma}^+} \equiv  \frac{1}{2}{\rm ln}\left(
\frac{4{\cal B}}{T^2 - T_{min}^2}\right)
\qquad {\tilde \sigma} ^- \equiv  - {\rm ln}(-x^-) 
\label{sigmas}
\ee
in terms of which the line element reads:
\be
ds^2=\frac{-2 {\cal F}({\tilde \sigma}^+) e^{2{\tilde \sigma} ^+ - {\tilde \sigma} ^-}
d{\tilde \sigma} ^+ d{\tilde \sigma} ^- }{A\frac{\lambda ^2}{2} 
{\rm ln}T_{min}^2/4 
+ \frac{4A (\lambda ^2/2) {\cal B}}{T_{min}^2}
e^{-e^{2{\tilde \sigma}^+}} + 2e^{-{\tilde \sigma}^-}}
\qquad A=4^{\lambda ^2/2}~a~C_1 
\label{dsA}
\ee
where
\be
{\cal F}({\tilde \sigma}^+)=(1 + T_{min}^2
{\rm exp}(e^{2{\tilde \sigma}^+})/4 {\cal B})^{-1}
\label{fromtext}
\ee 
This geometry has the following asymptotic behaviour
at early and late times:
\bea 
&~& ds^2_{in} =-e^{2{\tilde \sigma}^+}d{\tilde \sigma}^+ d{\tilde
\sigma }^-,
\qquad t \rightarrow -\infty \nn \\
&~& ds^2_{out} =-\frac{2{\cal F}({\tilde \sigma}^+)}{b + 2 e^{-{\tilde
\sigma}^-}}
e^{2{\tilde \sigma}^+ - {\tilde \sigma}^-}
d{\tilde \sigma}^+ d{\tilde \sigma }^-, \nn \\
&~&b \equiv A\frac{\lambda ^2}{2} {\rm ln}T_{min}^2/4 + \delta > 0 
\qquad t \rightarrow + \infty 
\label{inout}
\eea
where $\delta $ is related to the time cut-offs introduced earlier,
and is such that the signature of the metric does not change
as time flows from $-\Lambda $ to $\Lambda$. 
\pr
In the `in' vacuum $t \rightarrow -\infty $, the modes
in terms of which a spectator scalar field should be expanded are
standard Minkowski modes for right movers only, i.e., functions 
of $x^{-}$, as is
appropriate for an incoming wave. In this construction, the matter
field $T(x^{+})$ constitutes an infalling matter sector.  
The `in' modes are therefore:
\be
     u_{in} =\frac{1}{\sqrt{2\omega _1}}e^{-i\omega _1 \Sigma ^{-}}
\qquad \Sigma ^{+} =\frac{1}{2}e^{2{\tilde \sigma}^+} ~;~\Sigma ^- ={\tilde \sigma}^- 
\label{inmodes}
\ee
On the other hand, the `out' modes are defined by:
\be
     u_{out} =\frac{1}{\sqrt{2\omega _2}}e^{\pm i\omega _2
{\rm ln}(1 + \frac{2}{b}e^{-\Sigma ^-})}
\label{outmodes}
\ee
We notice immediately that the `out' spectrum contains both
positive and negative frequencies.
\pr
We focus on the positive frequency modes,
which determine the particle creation number for an external 
observer. This is determined as usual by the 
Bogolubov $\beta $ 
coefficient~\cite{birrel}, which in this case is easily found to be:
\bea 
&~&\beta _{\omega_1 \omega_2}^{>0}=
\frac{i}{\pi}\int _{-\infty}^0 d\Sigma ^- 
\frac{1}{\sqrt{2\omega _2}}e^{i\omega _2 {\rm ln}(1 + 
\frac{2}{b}e^{-\Sigma ^-})}
\partial _{\Sigma ^-}\left(\frac{1}{\sqrt{2\omega _1}}
e^{-i\omega_1 \Sigma ^-}\right) = \nn \\
&~&\frac{i}{2 \pi (\omega _1 + \omega _2 + i \epsilon )}
\sqrt {\frac{\omega _1}{\omega _2 + i \epsilon }}
\left(\frac{b}{2}\right)^{\epsilon - i(\omega _1 + \omega _2)}\nn \\
&~&_2F_1 (\epsilon - i(\omega _1 + \omega _2), \epsilon - i \omega _2, 
1  - i (\omega _1 + \omega _2); -\frac{b}{2})
\label{bogolubov}
\eea
where $\epsilon  > 0 $ specifies the infrared pole prescription for the
frequencies, whose form is dictated 
by the defining properties of the integral representation 
of the hypergeometric function. 
Notice that the presence of $\epsilon$ 
guarantees a smooth connection with the case $b =0$, corresponding to
the case with cosmological constant $\lambda =0$.
\pr
The corresponding particle number is given by;
\be
 N_{\omega _2} = \int _0^\infty d\omega _1 |\beta _{\omega _1\omega _2}|^2 
\label{creation}
\ee
The spectrum is clearly {\it not} thermal,
This can also 
as can be seen analytically from (\ref{bogolubov}),(\ref{creation})
in the case of $b << 1$, where we may expand
the hypergeometric function in 
powers of $b$.~\footnote{Note that the specific hypergeometric function 
$_2F_1(a,b,c;z)$
appearing in (\ref{bogolubov}) converges as a series in  $z$ 
in the entire circle of $|z| <1$.} 
For very small $\beta$ and low frequencies $\omega _2$, 
the leading terms in (\ref{creation}) have  
an inverse-power-law dependence on $\omega _2$,
\be
  N_{\omega _2} \sim \frac{1}{\omega _2}\int _0^\Lambda d\omega _1  
\frac{\omega _1}{(\omega _1 + \omega _2)^2} = 
\frac{1}{\omega _2}\left( {\rm ln}\frac{\Lambda + \omega _2}{\omega _2} 
- \frac{\Lambda }{\Lambda + \omega _2} + {\cal O}(b) \right ) 
\label{power}
\ee
and it can easily be
checked that the ${\cal O}(b)$ terms 
in the expansion of 
the hypergeometric function
are finite as $\Lambda \rightarrow \infty$. 
\pr
Further analytic support for a deviation from a thermal distribution
comes by representing the conformal factor $\rho$ 
(\ref{fullsol}) in the form: 
\bea
e^{-2\rho}= &~& C_1 - 4^{\frac{\lambda ^2}{2}}\frac{\lambda ^2 a}{2}
C_1 x^+ - \lambda ^2 x^+x^- 
+ \nn \\
&~& C_1F(1, 1 + \frac{\lambda ^2}{2}; -\frac{T^2}{4}) - C_1 
\label{rho}
\eea
where we have used properties of the incomplete $\Gamma $ function 
(\ref{completesolution}) for $\lambda^2 <<1 $ and $T << 1$.
The upper line in formula (\ref{rho}) coincides with the 
conformal factor corresponding to the static black hole 
of~\cite{nelson}, which has an ADM mass given by $\lambda C_1 >0$, 
and a horizon given by $x_H^-=-4^{\lambda ^2/2}(\frac{a}{2})C_1$. 
This static black hole is known~\cite{nelson} to produce a thermal 
radiation spectrum, as can readily be seen by the computation 
of the particle-creation number corresponding to the first
line of (\ref{rho}), which is obtained by integrating over 
the positive-frequency modes:
\be 
\beta _{\omega _1\omega _2}=\frac{1}{2\pi \lambda} 
\sqrt{\frac{\omega_1}{\omega_2 - i\epsilon}} 
\left(4^{\lambda ^2/2}(\frac{a}{2})C_1 \lambda \right)^{i\omega_2/\lambda}
B(-\frac{i}{\lambda}(\omega_1 + \omega _2  -i\epsilon), 1 +
\frac{i}{\lambda}\omega_2)
\label{betanelson}
\ee
where $\epsilon$ is the infrared cut-off.
The corresponding particle creation 
number (\ref{creation}) has a thermal spectrum with temperature 
$T=\lambda/2\pi$. 
\pr
We have already commented that the spectrum calculated from the
full version of (\ref{rho}) is not thermal. Fig.~2 displays
numerically the ratio of 
the thermal spectrum 
calculated from the first line of (\ref{rho}) divided by the  
full spectrum (\ref{creation},\ref{bogolubov}).  
It is evident that the $\lambda \ne 0$ case is certainly {\it not}
a stable equilibrium state.
\pr
We have considered above the case of  a 
spectator particle moving in the background of the metric
(\ref{fullsol}), with the particle 
in a Minkowski space time in the asymptotic past `before' the formation of
the black hole. Similar conclusions are reached by considering a related 
problem, in which
the propagation of the spectator particle starts
{\it not} in the flat Minkowski vacuum,
but in an asymptotic region of the black-hole solution 
(\ref{fullsol}). In this case, one cannot reach the intitial 
singularity shown in Fig.~1, but instead imposes a cut-off ${\cal B}$ 
in the denominator of the line element of the corresponding `in' states:
\bea 
&~&ds_{in}^2=-\frac{2{\cal F}({\tilde \sigma}^+)~e^{2{\tilde
\sigma}^+} 
d{\tilde \sigma }^+d{\tilde \sigma}^-}{{\cal B} + 2e^{-{\tilde \sigma}^-}}, 
\nn \\
&~&ds_{out}^2=-\frac{2{\cal F}({\tilde \sigma}^+)~e^{2{\tilde
\sigma}^+} d{\tilde \sigma }^+
d{\tilde \sigma}^-}
{b + 2e^{-{\tilde \sigma}^-}}, 
\label{out}
\eea
where ${\cal B} > b$, $b$ and ${\cal F}({\tilde \sigma}^+)$ 
have been defined previously, 
and their relation ensures that matter can be treated in the
weak-field approximation.
\pr
Following an analysis similar to the one leading to (\ref{bogolubov}),
one obtains the Bogolubov $\beta _{\omega_1\omega_2}$ coefficient
for the non-trivial vacuum `seen' by an observer at future infinity:
\bea
&~&\beta '_{\omega_1\omega_2} =
\frac{i}{\pi}\int _{-\infty}^0 d\Sigma ^- 
\frac{1}{\sqrt{2\omega _2}}e^{i\omega _2 {\rm ln}[1 + \frac{{\cal B}}{b}
(e^{-\Sigma ^-}-1)]}
\partial _{\Sigma ^-}\left(\frac{1}{\sqrt{2\omega _1}}
e^{-i\omega_1 \Sigma ^-}\right) = \nn \\
&~&\frac{i}{2 \pi (\omega _1 + \omega _2 + i \epsilon )}
\sqrt {\frac{\omega _1}{\omega _2 + i \epsilon }}
\left(\frac{b}{{\cal B}}\right)^{\epsilon - i \omega _2}\nn \\
&~&_2F_1 (\epsilon - i\omega _2, \epsilon - i (\omega _1 + \omega _2),
1  - i (\omega _1 + \omega _2); 1-\frac{b}{{\cal B}})
\label{newbog}
\eea
We plot in Fig.~3 the ratio of the particle-creation
number in a thermal distribution to that corresponding to 
(\ref{newbog}).  
As in the previous analysis, we find that the
particle-number distribution is {\it non-thermal} when the
cosmological constant $\lambda \ne 0$.
Moreover, by taking the limit
$b \rightarrow 0$, such that $b < {\cal B}$, a smooth connection 
with the stable case $\lambda = 0$ is obtained.

\section{Renormalization-Group Flow of $\lambda \rightarrow 0$}

It is tempting to guess that the equilibrium case $\lambda = 0$
constitutes the end-point of the time evolution
of the $\lambda^2 \ne 0$ solutions, which we have shown to be
unstable, as exemplified by the non-thermal particle creation
seen in (\ref{bogolubov}) and (\ref{newbog}). As we now
demonstrate, this point of view is supported by a
Liouville-string interpretation of the space-time metric
(\ref{rindler}).  
\pr
Let us review briefly the identification of the world-sheet Liouville 
field as target time~\cite{emn,kogan}, before proceeding to this
demonstration. 
Consider a world-sheet $\sigma$-model action in a non-conformal
background $g^i \equiv 
\{ G_{MN},\Phi,T \} $, described by the
coordinates $\{ X^M \}$,$M=1, \dots  D-1$:
\be 
S=\frac{1}{4\pi \alpha '} \int _\Sigma d^2z G_{MN} 
\partial X^M {\overline X}^N  
-2\int _\Sigma d^2z \Phi (X) R^{(2)} d^2z + 
\int _\Sigma d^2z T(X) 
\label{sigma} 
\ee
The conformal invariance of
this $\sigma$ model can be restored by the dynamics of the Liouville
field~\cite{ddk},
which can, according to the analysis of~\cite{emn}, be viewed as 
a local renormalization-group scale on the world-sheet, 
$\varphi (z, {\bar z})$, as is appropriate for
renormalization in a curved space~\cite{osborn}. 
Within this approach, one must add new counterterms 
to the $\sigma$-model 
action (\ref{sigma}), which are known from power counting to assume the
generic form: 
\be 
    S_{ct}[g] = \int _\Sigma  d^2z \left( 
\partial g^i {\cal G}_{ij} {\overline \partial} g^j 
+ \Lambda [g] + {\cal F} [g] R^{(2)} \right)  
\label{counterterms} 
\ee
where ${\cal G}_{ij}$ is related to the Zamolodchikov 
metric in coupling-constant space~\cite{zam},
and the indices $i$ run over background fields 
$G_{MN}, \Phi, T$.~\footnote{We exhibit here only the basic
structure of the various counterterms. 
For details we refer the reader to the literature~\cite{emn,osborn}.}
We observe that the world-sheet dependence of the renormalized 
coupling constants $g^i$ on the local Liouville renormalization scale
$\rho $ in the first counterterm in (\ref{counterterms}) 
yields a kinetic term for the Liouville mode~\cite{emn}:
\be
\int _\Sigma d^2z 
(\partial \rho)\beta^i {\cal G}_{ij} \beta^j {\overline \partial} \rho
\label{Lkin}
\ee
where the $\beta ^i$ are the world-sheet renormalization-group $\beta$
functions, which can be regarded as
Weyl anomaly coefficients~\cite{osborn,emn}.            
In addition, as discussed in \cite{ddk},
one may parametrize the world-sheet metric 
as $\gamma _{\alpha\beta}=e^{\rho}{\hat \gamma}_{\alpha\beta}$, which implies
additional Liouville terms in the effective renormalized $\sigma$-model:
\be 
-\frac{1}{48\pi}[\partial \rho {\overline \partial} \rho 
+ \rho R^{(2)} + {\rm counter~terms}]
\label{liouville}
\ee
Combining (\ref{liouville}) and (\ref{counterterms}) 
we see a term linear in the Liouville scale $\rho$
in the renormalized dilaton field 
\be
\Phi _R = \Phi (X) + {\cal F}[g] + \rho 
\label{dilback}
\ee
Moreover, it has been shown~\cite{emn,ddk} that
world-sheet kinetic terms for the $\rho$ field have the form: 
$-(C[g]-26)\partial \rho {\overline \partial} \rho$, with a
{\it negative} sign
relative to the $X^M$ kinetric terms, where $C[g]$ is the 
field-dependent effective central 
charge of the non-conformal initial theory. 
This implies 
that for supercritical strings~\cite{aben} with $C > 26$ the Liouville 
field acquires a Minkowskian time signature, which led 
the authors of~\cite{emn,aben,kogan} to interpet it as 
target time. 
\pr
Within this interpretation 
of the Liouville field as target time, 
the $\sigma$ model in the $D$-dimensional space time 
$(\rho, X^M)$ is {\it conformal}. 
{}From the above discussion and (\ref{dilback}),  
it is clear that 
the  conformal solution (\ref{rindler}), (\ref{dilatonback3}) 
falls in the above category, once we
identify the conformal scale factor of the metric 
$\rho$ with the Liouville field as above.
The thermodynamic instability of the $\lambda \ne 0$ model,
discussed above, suggests a renormalization-group
flow towards the solution $\lambda ^2 =0$.
This is consistent with the fact that
(\ref{lamb}) corresponds to the minimum, critical value 
of the central charge, in agreement with the Zamolodchikov $c$ theorem
for renormalization-group flow in unitary $\sigma$ models. 
The flow of the system from an initial unstable point 
is triggered by matter deformations, and thus 
the entire phenomenon described in this article may be considered
as reflecting the back reaction of matter on the space-time geometry.
The fact that this back reaction is present in the classical 
target-space solution is not in contradiction with the 
quantum nature of the phenomenon in the Liouville $\sigma$-model
language, since the latter always 
describes a conformal string in $(\rho =t, X)$ space time.
\pr
{}From this point of view, the rate of flow of the system towards the
flat space time 
with vanishing cosmological constant is 
computable using Liouville dynamics~\cite{emncosmol}. 
The rate of change of the cosmological constant 
during the non-equilibrium phase is
obtained from the Zamolodchikov $c$ theorem~\cite{zam}, appropriately
extended 
to Liouville strings, once one identifies the Liouville 
field with a local renormalization scale on the world-sheet~\cite{emn}:
\be
\frac{ \partial}{\partial t} \lambda ^2  \sim - \sum _{i,j}~\beta ^i <V_iV_j> \beta^j 
\label{ctheorem}
\ee
where the $V_i$ are the 
vertex operators corresponding to deformations $g^i$, and the summation
over $i, j$ also includes
spatial integrations $\int dr \int dr ' $.
The two-point correlators $<V_i V_j>$ constitute 
a metric in theory space~\cite{zam}, which is positive definite 
for unitary world-sheet theories, 
implying an irreversible time flow~\cite{emn}.
\pr
Within this approach, 
it is immediate to deduce from (\ref{ctheorem}) 
the time dependence of $\lambda (t)$.
In the case 
of small $\lambda$ and weak matter fields $T$, the leading-order 
effect is associated with the graviton $\beta$ functions.
To ${\cal O}(\alpha ')$, the latter are  
proportional to $G_{MN}~R$, where $R$ is the scalar curvature
given in~(\ref{curvature}).
Using the fact that, to this order, the
Zamolodchikov  metric $<V_iV_j>$ is proportional to 
$\delta _{ij}\delta (r - r') $, one finds:
\be
\frac{\partial}{\partial t} \lambda ^2 \sim - \int dr R^2(r,t) 
\simeq - A~e^t~\lambda ^4  + {\cal O}(\lambda ^4 T^2, \lambda ^6)
\label{rge}
\ee
where 
\be
A=\left (4C_1^2 a^2 
\int _0^\infty dr e^{-r} \right)= 4C_1^2 a^2 
\label{ctheorem2}
\ee
and we have used
the transformation (\ref{transform})
to pass from the light-cone variables to $(r, t)$. 
The solution of (\ref{rge}) and (\ref{ctheorem2}), with the initial
condition $\lambda ^2 (0)=\lambda _0 ^2 << 1$, is: 
\be
       \lambda ^2 (t) = \frac{\lambda _0^2 }{1 + \lambda _0^2 A (e^t - 1)}
\label{solution}
\ee
Thus we have a quantitative description of the flow in Liouville time:
$\lambda ^2 \rightarrow 0$ as $t \rightarrow \infty$.

\begin{centering}
\begin{figure}[t]
\epsfxsize=4in
\centerline{\epsffile{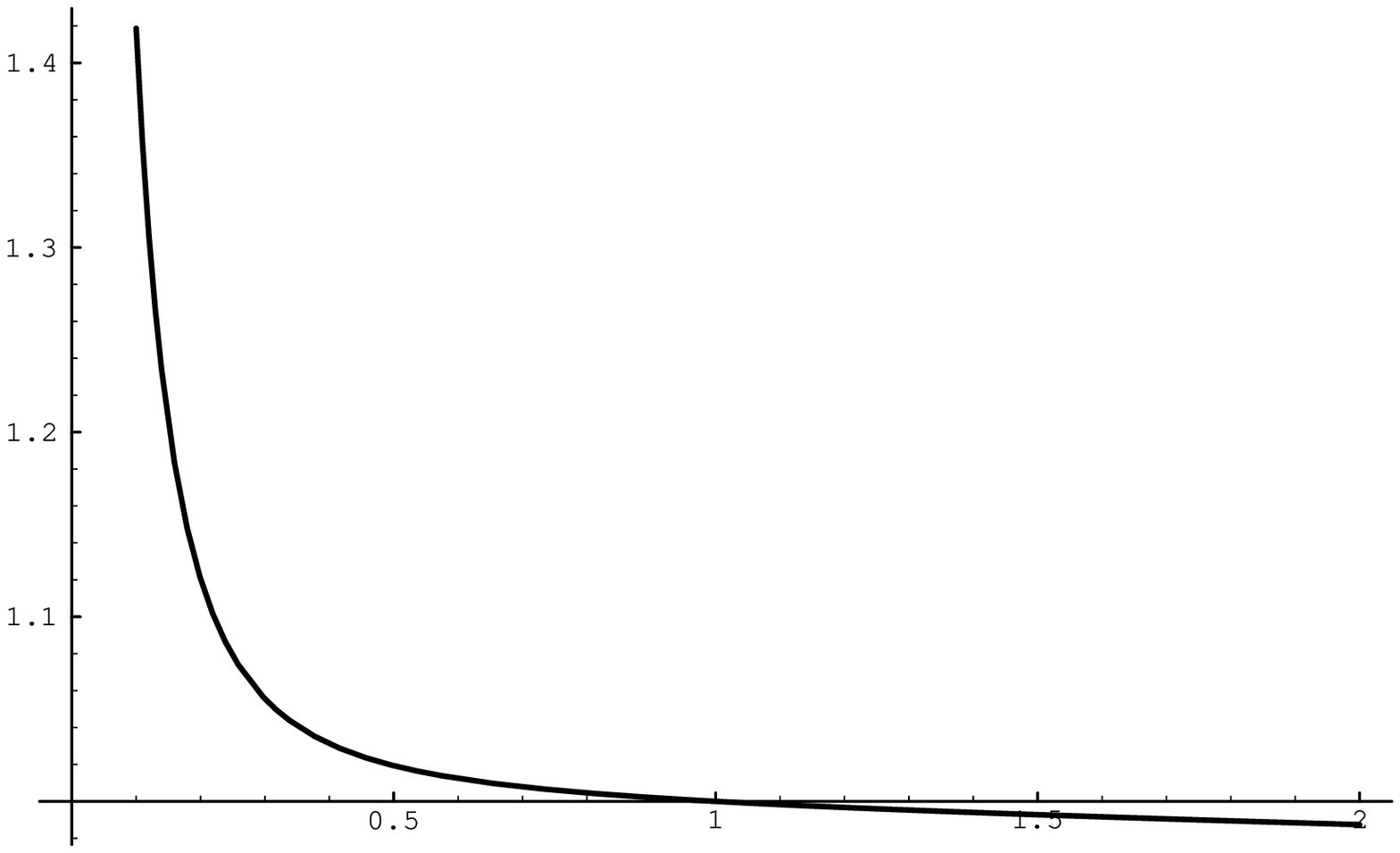}}
\vspace{1cm}
\caption{{\it Plot of the ratio of the particle number corresponding to a 
thermal distribution to that 
corresponding to the Bogolubov coefficient (\ref{bogolubov}).}}
\label{fig2}
\end{figure}
\end{centering}

\begin{centering}
\begin{figure}[t]
\epsfxsize=4in
\centerline{\epsffile{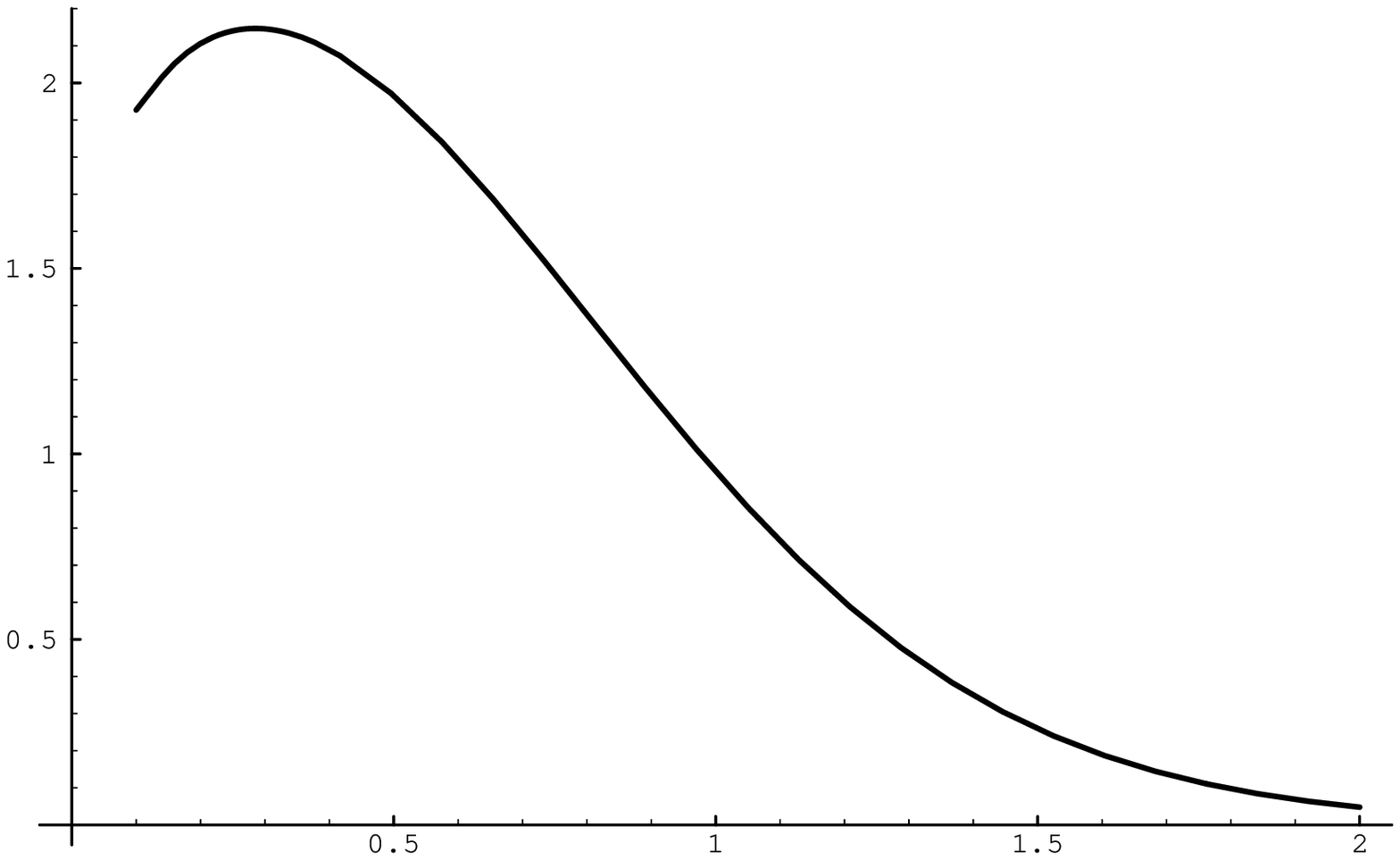}}
\vspace{1cm}
\caption{{\it Plot of the ratio of the particle number corresponding to a 
thermal distribution to that 
corresponding to the Bogolubov coefficient (\ref{newbog}).}} 
\label{fig3}
\end{figure}
\end{centering}

\section{Conclusions}

We have discussed in this paper the physical interpretation
of the two-dimensional dilaton-matter black-hole solutions
found in~\cite{tachyon}. The case with zero
cosmological constant $\lambda = 0$ is a stable, equilibium configuration,
whereas the $\lambda \ne 0$ solutions exhibit non-equilibrium, non-thermal
particle production. This suggests that matter back reaction relaxes
the comsological constant to zero. Indeed, we have demonstrated this to be
the case, by identifying the world-sheet Liouville field with target
time, and using the renormalization-group equation of the world-sheet
$\sigma$ model to calculate the rate of this irreversible flow. Much
work remains to be done along this line, but the results presented here
demonstrate that the two-dimensional model of~\cite{tachyon} may serve as a
useful model for studying gravitational dynamics, and as a laboratory
for developing the Liouville description of time~\cite{emn}.

\section*{Acknowledgements} 

The work of N.E.M. is supported by a P.P.A.R.C. Advanced Fellowship, and
the work of D.V.N. is supported in part by D.O.E. Grant
DEFG05-91-GR-40633.

\end{document}